\documentclass{article}

\usepackage{amssymb,amsmath}
\usepackage{graphicx}
\usepackage{cite}

\newcommand{\Real}{\mathbb{R}}

\newcommand{\Y}{\mathbf{Y}}
\newcommand{\y}{\mathbf{y}}
\newcommand{\Q}{\mathbf{Q}}
\newcommand{\X}{\mathbf{X}}
\newcommand{\Z}{\mathbf{Z}}
\newcommand{\W}{\mathbf{W}}
\newcommand{\F}{\mathbf{F}}
\newcommand{\f}{\mathbf{f}}
\newcommand{\A}{\mathbf{A}}

\newcommand{\G}{\mathbf{G}}

\newcommand{\R}{\mathbf{R}}
\newcommand{\V}{\mathbf{V}}
\newcommand{\U}{\mathbf{U}}
\newcommand{\id}{\mathbf{I}}

\newcommand{\zero}{\mathbf{0}}
\newcommand{\position}{\mathbf{r}}
\newcommand{\Position}{\mathbf{R}}
\newcommand{\projector}{\mathbf{P}}

\newcommand{\superX}{\mathbf{\Xi}}
\newcommand{\superF}{\mathbf{\Phi}}

\newcommand{\extpotential}{\mathbf{v}_\mathrm{ext}}
\newcommand{\intpotential}{\mathbf{v}_\mathrm{int}}
\newcommand{\laplace}{\mathbf{L}}
\newcommand{\density}{\mathbf{n}}
\newcommand{\diag}{\mathrm{diag}}

\newcommand{\electrons}{n_e}

\newcommand{\step}{\tau}
\newcommand{\transport}{\mathbf{T}}

\newcommand{\tangentspace}[1]{\mathcal{T}_{#1}\mathcal{M}}

\newcommand{\manifold}{\mathcal{M}}

\title{Direct Minimization for Ensemble Electronic Structure Calculations}

\author{K. Baarman\footnote{Department of Mathematics and Systems Analysis, Aalto University School of Science, Espoo, Finland, e-mail: \texttt{kurt.baarman@aalto.fi}}, V. Havu\footnote{Department of Applied Physics, Aalto University School of Science, Espoo, Finland}, and T. Eirola\(^*\)}

\begin{document}
\maketitle

\begin{abstract}
We consider a direct optimization approach for ensemble density functional theory electronic structure calculations.
The update operator for the electronic orbitals takes the structure of the Stiefel manifold into account and we present an optimization scheme for the occupation numbers that ensures that the constraints remain satisfied.
We also compare sequential and simultaneous quasi-Newton and nonlinear conjugate gradient optimization procedures, and demonstrate that simultaneous optimization of the electronic orbitals and occupation numbers improve performance compared to the sequential approach.
\end{abstract}

%\keywords{Quasi-Newton method \and nonlinear conjugate gradient \and ensemble optimization \and electronic structure \and Stiefel manifold}

%\begin{AMS}
%49M15, 65K10, 65Z05
%\end{AMS}

\section{Introduction}

Advances in computer power and numerical methods during the past few decades has dramatically increased the scope of electronic structure problems that can be computationally studied.
Kohn-Sham density functional theory (DFT) methods can be used to reach precision comparable to experimental accuracy for insulators and semiconductors, while metallic systems remain more challenging.
Metallic systems lack a gap between occupied and unoccupied electronic states in the energy spectrum, which leads to slower convergence compared to insulators and semiconductors.
Smearing of the Fermi surface is often used to enable convergence of metallic systems as well as insulators at positive temperatures.
Ensemble DFT permits direct computation of the occupation numbers of the orbitals based on the entropic term in the Helmholtz free energy.
We consider an optimization problem where the target functional \(A\) corresponds to the Helmholtz free energy and the variables \(\X\) and \(\f\) to the electronic orbitals and occupation numbers respectively.

The optimization problem is therefore
\begin{equation}\label{eq:problem}
\mathrm{minimize}\;A(\X,\f),
\end{equation}
subject to
\begin{equation}
\X\in\manifold=\{\X\in\Real^{m\times n}\,|\,\X^T\X = \id\}.
\end{equation}
Furthermore \(\f\in\Real^n\) with \(\sum_{i=1}^n f_i = \electrons\) and \(0 \leq f_i \leq 1\), where \(\electrons\in\mathbb{N}\) is the number of electrons in the system and \(\electrons \leq n\).
We also assume that \(\nabla_\X A(\X,\f)\) and \(\nabla_\f A(\X,\f)\) are available, but expensive to compute.
However, due to the form of \(A(\X,\f)\) the price to compute \(A\), \(\nabla_\X A\), and \(\nabla_\f A\) simultaneously is comparable to computing one of them separately.
%, and we observe that the orbital gradient has the form
%\begin{equation}\label{eq:gradform}
%\nabla_\X A = \mathbf{H}(\X,\f)\,\X\,\diag(\f),
%\end{equation}
%as the unoccupied electronic orbitals do not contribute to the energy of the system.
Furthermore we assume that \(m \gg n\), and that \(m\) is sufficiently large as to make storage of and operation with full \(m\times m\) matrices prohibitively expensive.

The orthogonality constraint on \(\X\) means that \(\manifold\subset\Real^{m \times n}\) defines the Stiefel manifold, which has the tangent space
\begin{equation}
\tangentspace{\X} = \{\Y = \X\A+\Z\;|\;\A^T=-\A\;\mathrm{and}\;\Z^T\X=\zero\},
\end{equation}
where \(\Y, \Z \in \Real^{m\times n}\) and \(\A\in\Real^{n\times n}\).
We use the standard inner product
\begin{equation}
(\X,\Y) = \mathrm{trace}(\X^T\Y).
\end{equation}
Given an arbitrary matrix \(\W\in\Real^{m\times n}\) we can orthogonally project it onto \(\tangentspace{\X}\) with
\begin{equation}
\Y = \projector_\X(\W) = (\id - \tfrac{1}{2}\X\X^T)\W - \tfrac{1}{2}\X\W^T\X.
\end{equation}

Minimization approaches to non-temperature dependent DFT do not in general permit fractional occupation of electronic orbitals \cite{kresse_metal,voorhis2002,vandevondele2003,saad,zhou2006b,bekas2008}.
In contrast, explicit minimization with regards to occupation numbers permits fractional occupation based on the entropy functional of the Helmholtz free energy and can improve convergence, especially for metallic systems \cite{marzari,cances2001,freysoldt}.
It is also possible to transform Equation~\eqref{eq:problem} into a nonlinear eigenvalue problem that can be solved through a self consistent field iteration \cite{pulay,kresse_metal,luke,saad}.
%\marginpar{I didn't understand this strikethrough. Depending on where we send it I'll check that the cites is on the correct side of the dots for the style.}
The absence of well separated occupied and unoccupied orbitals make metallic systems challenging to compute, and broadening of the Fermi surface is used to facilitate convergence \cite{Wagner1998,Blum2009}.
This broadening is often achieved by assigning the orbitals close to the Fermi level a fractional occupation number determined by the energy of the electronic orbital \cite{Mermin1965,Fu1983,Methfessel1989}.
Direct minimization on the other hand does not require the orbital energies to be computed at every step, and these broadening schemes are therefore not well suited for minimization methods.

In \cite{edelman1998} a framework for optimization methods on the Stiefel and Grassmann manifolds is presented, while~\cite{chu1983} discusses a Newton-like iteration scheme on a more general manifold.
Univariate optimization methods for the Stiefel manifold is presented in \cite{celledoni2008}, where identity plus rank one Householder transforms are given as one possible choice for moving on the manifold.
The choice of coordinates can also be based on a QR factorization and polar decompositions~\cite{celledoni2002,dieci2003} or Lie groups~\cite{krogstad2003}.
An overview of geometric numerical integration techniques can be found in~\cite{lubich_hairer_wanner}.

In Section~\ref{sec:optimization} we first recall the nonlinear conjugate gradient and the quasi-Newton methods adapted for use on the Stiefel manifold. We then present an optimization procedure for the occupation numbers and end the section by presenting a simultaneous orbital-occupation optimization strategies.
Then, in Section~\ref{sec:numerics} we numerically demonstrate the method on a model problem that includes nonlinearities similar to a DFT problem.
The conclusions are finally presented in Section~\ref{sec:conclusion}.

\section{Optimization with orthogonality constraints}\label{sec:optimization}

\subsection{Update and transport}\label{sec:update}

We ensure that \(\X_{k+1}\) satisfies the orthogonality constraint by using a unitary update operator \(\U\) which maps \(\manifold\rightarrow\manifold\).
%~\cite{baarman2012a}.
%In other words, \(\X_{k+1} = \U\X_k\in\manifold\) whenever \(\X_k\in\manifold\).
A search direction \(\Y\in\tangentspace{\X}\) given by an optimization procedure can be written
\begin{equation}\label{eq:decomp}
\Y = \X\A+\Q\R,
\end{equation}
where \(\Q\in\Real^{m\times n}\), \(\A,\R\in\Real^{n\times n}\), \(\A^T=-\A\), \(\Q^T\Q = \id\), and \(\Q^T\X=\zero\).
%It is not necessary that \(\Q\R\) is constructed by the QR~decomposition as long as \(\Q\) satisfies the requirements placed on it.
If the terms in Equation~\eqref{eq:decomp} are not full rank the size of the matrices can be adjusted accordingly.

%If we follow \(\Y\) to update \(\X\) along a Stiefel geodesic we obtain the update operator for \(\X\)~\cite[p. 326]{edelman1998}
If we follow \(\Y\) to update \(\X\) along a Stiefel geodesic we obtain the update operator for \(\X\)~\cite{edelman1998}
\begin{equation}
\U=\begin{bmatrix}\X&\Q\end{bmatrix}\exp\left(\step\begin{bmatrix}\A & -\R^T\\\R&\zero\end{bmatrix}\right)\begin{bmatrix}\id&\zero\end{bmatrix}^T,
\end{equation}
with step length parameter \(\step\).
The update operator generalized for an arbitrary matrix in span\((\X,\Q)\) is
\begin{equation}\label{eq:unitary_update}
\U=\begin{bmatrix}\X&\Q\end{bmatrix}\exp\left(\step\begin{bmatrix}\A & -\R^T\\\R&\zero\end{bmatrix}\right)\begin{bmatrix}\X&\Q\end{bmatrix}^T,
\end{equation}
where the orthogonality of \(\X\) and \(\Q\) has been exploited.

In order to use information gained from previous evaluations of \(A\) and \(\nabla A\) we must take \(\manifold\) into account.
This requires us to transport vectors \(\Y\in\tangentspace{\X}\) to \(\tangentspace{\U\X}\) with the transport operator
\begin{equation}
\transport = \id_m+\begin{bmatrix}\X&\Q\end{bmatrix}\left(\exp\left(\step\begin{bmatrix}\A & -\R^T\\\R&\zero\end{bmatrix}\right)-\id_{2n}\right)\begin{bmatrix}\X&\Q\end{bmatrix}^T.
\end{equation}
Here \(\id_m\in\Real^{m\times m}\) and \(\id_{2n}\in\Real^{2n\times 2n}\), and \(\transport\) does not modify matrices \(\Z\) that satisfy \(\begin{bmatrix}\X&\Q\end{bmatrix}^T\Z=\zero\).

\textbf{Remark 1:} The closely related Grassmann manifold is identical to the Stiefel manifold with the addition of the homogeneity condition \(A(\X) = A(\X\Q)\), where \(\Q\) is orthogonal.
The homogeneity condition is satisfied for orbitals with identical occupation numbers, but does not generally hold for ensemble DFT.
A discussion of direct minimization with integer occupation numbers is presented in~\cite{baarman2012a}.

\subsection{Nonlinear conjugate gradients}\label{sec:nlcg}
The conjugate gradient (CG) method can be viewed as an optimization method for a quadratic problem. Several generalizations of the CG method have been presented to solve optimization problems that are not quadratic \cite{nocedal}.
Below, we review a nonlinear CG (NLCG) method adapted to account for the curvature of the manifold \cite{edelman1998}.

Given \(\X_0\) which satisfies \(\X_0^T\X_0 = \id\), the gradient projected onto \(\tangentspace{\X_0}\) is
\begin{equation}
\F_0 = \projector_{\X_0}(\nabla_\X A(\X_0,\f_0)),
\end{equation}
and the initial search direction is the direction of steepest descent
\begin{equation}
\Y_0 = -\F_0.
\end{equation}

On the manifold the NLCG method then proceeds by minimizing \(A\) along the path defined by the search direction \(\Y_k\). In practice we evaluate \(A\) once along the search direction and construct a quadratic approximation that we minimize.
The step length, \(\step_k\), that minimizes \(A\) along the search direction is then used to update \(\X_k\) such that
\begin{equation}
\X_{k+1} = \transport(\step_k)\X_k,
\end{equation}
and the gradient and search directions are transported to \(\tangentspace{\X_{k+1}}\) by \(\transport(\step_k)\).
The new projected gradient
\begin{equation}\label{eq:gradient}
\F_{k+1} = \projector_{\X_{k+1}}(\nabla_\X A(\X_{k+1},\f_{k+1})),
\end{equation}
and search direction
\begin{equation}
\Y_{k+1} = -\F_{k+1} + \gamma_k \transport (\step_k) \Y_k,
\end{equation}
are then computed where
\begin{equation}
\gamma_k = \frac{(\F_{k+1}-\transport(\step_k)\F_k,\F_{k+1})}{(\F_k,\F_k)}.
\end{equation}

The step length is determined by the minimizer of a quadratic approximation of \(A\) along the search direction.
The quadratic approximation is constructed by taking a trial step length \(\step_e = \tfrac{1}{10}\max(\step_\mathrm{min},\step_{k-1})\), where \(\step_\mathrm{min}\) is a predefined minimum trial step length and computing
\begin{align}\label{eq:step}
p(0) &= A(\X,\f),\nonumber\\
p(\step_e) &= A(\transport(\step_e)\X,\f),\\
p'(0) &= (\Y,\nabla_\X A(\X,\f)).\nonumber
\end{align}
Then solve \(\step_k\) and limit it by \(2\step_{k-1}\), and construct the update \(\transport(\step_k)\).
This approximate line search requires one extra evaluation of \(A\) per step.

\subsection{Quasi-Newton method}
The quasi-Newton (QN) method is similar to Newton's method, but replaces the inverse Hessian with an approximation.
This is frequently possible even when the Hessian is not available, and can still be used to improve performance for a badly conditioned minimization problem.

We base the QN method on Broyden's second or \emph{bad} generalized update to construct the approximate inverse Hessian, \(\G\), of \(A\) at \(\X_k\). While Broyden's second update does not construct a symmetric approximation, or ensure that the approximation is positive definite it is a robust update choice for electronic structure calculations \cite{luke,baarman2011a}.
Furthermore, \(\X\) and \(\nabla_\X A\) are \(\Real^{m\times n}\) matrices, which we take into account when constructing the generalized Broyden update.
The secant condition is then
\begin{equation}\label{eq:secant}
\G \Delta \superF = \Delta \superX,
\end{equation}
where \(\Delta\superF\) and \(\Delta\superX\) are the collected orbital gradient and position differences projected onto the tangent space and transported to \(\tangentspace{\X_k}\). That is
\begin{equation}
\Delta\superF = \begin{bmatrix}\Delta\F_{k-1} & \transport(\step_{k-1})\Delta\F_{k-2} &\ldots& \transport(\step_{k-1})\ldots\transport(\step_{l+1})\Delta\F_{l}\end{bmatrix},
\end{equation}
and
\begin{equation}
\Delta\superX = \begin{bmatrix}\Delta\X_{k-1} & \transport(\step_{k-1})\Delta\X_{k-2} &\ldots& \transport(\step_{k-1})\ldots\transport(\step_{l+1})\Delta\X_{l}\end{bmatrix},
\end{equation}
for history length \(k-l\).
Here the gradient differences projected onto \(\tangentspace{\X_{i+1}}\) are
\begin{equation}
\Delta \F_i = \F_{i+1} - \transport(\step_i)\F_i,
\end{equation}
and \(\F_i\) is like in~\eqref{eq:gradient},
\begin{equation}
\F_{i} = \projector_{\X_{i}}(\nabla_\X A(\X_{i},\f_{i})).
\end{equation}
The projected occupation weighted orbital differences are
\begin{equation}
\Delta\X_i = \projector_{\X_{i+1}}\big(\X_{i+1}\,\diag(\f_{i+1})-\X_i\,\diag(\f_i)\big),
\end{equation}
and the motivation for including the weight is that the unoccupied electronic orbitals do not contribute to the energy of the system.
The no change condition is now
\begin{equation}
\Z = \G\Z\quad \forall\,\Z\;\;\mathrm{such\;that}\;\;\Z^T\Delta \superF = \zero.
\end{equation}
%\marginpar{Fang and Saad use group, but I tried to reformulate this.}
The secant and no change condition together correspond to the generalized Broyden's second update where all single orbital secant conditions are simultaneously enforced for the entire history length.
We can therefore use the generalized update formula \cite{fang}
\begin{equation}
\G = \mu\id + (\Delta \superX - \mu \Delta \superF) (\Delta \superF^T\Delta\superF)^{-1}\Delta\superF^T,
\end{equation}
where dropping the empty orbitals ensure that \(\Delta\superF^T\Delta\superF\) is nonsingular in practice.
%This reduces to the method of steepest descent (SD) with line search scaled by \(\mu\) when no history is included.
The search direction given by the QN method is then
\begin{equation}
\Y = -\G\F,
\end{equation}
and
\begin{equation}
\X_{k+1} = \U(\step_k)\X_k,
\end{equation}
where \(\Y\) determines \(\U\) as in Section~\eqref{sec:update}.
The line search is identical to the one described for the NLCG method in Section~\ref{sec:nlcg} with the addition of the constant underrelaxation \(\beta_\X\in\;]0,1]\) that we have included in the step length \(\step_k\).

In practice only the last few history steps contribute significantly to the rate of convergence.
Consequently, we discard the oldest trial solutions and gradient information when a predetermined history length is reached.%and reindex the remaining trial solutions and corresponding gradient information.

\subsection{Optimization of occupation numbers}

Given a set of electronic orbitals \(\X\) it is possible to further reduce \(A\) by optimizing \(\f\). Forcing occupation towards a uniform distribution increases contributions to \(A\) from higher energy states, while simultaneously increasing the entropy which contributes to a reduction of \(A\) at nonzero temperatures.
The relative strength of both of these effects determine the ground state of the system, and can lead to nonzero occupation of higher energy states at positive temperatures or due to nonlinear effects.

Therefore, given \(\X\), we want to find \(\f\) that minimizes \(A\). To keep the number of particles constant we determine the search direction \(\y\) which is the vector closest \(-\nabla_\f A(\X,\f)\) that ensures that the conditions \(\sum_{i=1}^n f_i = \electrons\) and \(0 \leq f_i \leq 1\) remain satisfied. To this end we solve
\begin{equation}\label{eq:occupation_direction}
\mathrm{minimize}\; \|\y + \nabla_\f A(\X,\f)\|,
\end{equation}
with the constraints \(\sum_{i=1}^n y_i = 0\), \(y_i \leq 0\) if \(f_i = 1\), and \(y_i \geq 0\) if \(f_i = 0\).
%\marginpar{The sum should actually be zero because we solve the search direction. I added an explanation, do we need an explicit warning?}
The first constraint on \(\y\) ensures that the minimization step conserves electrons while the second and third condition prohibits unphysical occupation numbers.
In practice we use the {\tt quadprog} routine available in MATLAB to solve this problem.
Given the search direction \(\y\) we minimize \(A\) by constructing a quadratic approximation similar to~\eqref{eq:step}.

After we have solved \(\y\) the occupation step length \(\sigma_k\) is determined like in Section~\ref{sec:nlcg} with the addition of the constant underrelaxation \(\beta_\f \in\;]0,1]\) included in \(\sigma_k\).
In addition, we ensure that the occupation remains physical by limiting \(\sigma_k\) with \(\sigma_{\mathrm{M}}\) such that \(0 \leq f_i+\sigma_{\mathrm{M}}y_i \leq 1\) for all \(i\).% where the subscript in this instance indicates entries in \(\f\) and \(\y\).
It is possible to take a longer step than \(\sigma_\mathrm{M}\) by recomputing \(\y\) from Equation~\eqref{eq:occupation_direction} with the updated boundary information when an entry in \(\f\) reaches the boundary of physical occupation, \(0\) or \(1\). However, convergence of occupation numbers is faster than orbital convergence, and the numbers of steps needed for convergence is therefore determined by the orbital convergence.
Furthermore, if the occupation number of the least populated orbital has been less than \(10^{-12}\) on two consecutive iterations we drop the associated orbitals.

\subsection{Simultaneous step size selection}

Typically an ensemble DFT problem is solved by sequentially optimizing the orbitals with fixed occupation numbers and then fixing the orbitals and optimizing the occupation numbers. This process is then repeated until a satisfactory solution is obtained.

The cost of evaluating \(A\), \(\nabla_\X A\), and \(\nabla_\f A\) is comparable to evaluating one of them separately, and simultaneous optimization of \(A\) with respect to \(\X\) and \(\f\) can for this reason potentially reduce computational effort.

Given a pair of search directions \((\Y\), \(\y)\) for the orbitals and occupation numbers respectively and starting guesses for step lengths, \(\step_{k-1}\) and \(\sigma_{k-1}\) we evaluate \(A\) and its gradients with the following trial step lengths
\begin{equation}
\step_e = \tfrac{1}{10}\max(\step_\mathrm{min},\step_{k-1}),
\end{equation}
and
\begin{equation}
\sigma_e = \mathrm{min}(\sigma_\mathrm{M},\tfrac{1}{10}\max(\sigma_\mathrm{min},\sigma_{k-1})).
\end{equation}
Here \(\step_\mathrm{min}\) and \(\sigma_\mathrm{min}\) are minimum trial step lengths.
With this we construct a quadratic surface approximation
\begin{equation}\label{eq:surface}
p(\step, \sigma) = c_1\step^2+c_2\sigma^2+c_3\step+c_4\sigma+c_5,
\end{equation}
that we use to simultaneously update both \(\X\) and \(\f\) by evaluation in one trial point. This surface is determined by the system of equations
\begin{align}\label{eq:surfsystem}
p(0,0) &= A(\X,\f),\nonumber\\
p_\step(0,0) &= (\Y,\nabla_\X A(\X,\f)), \nonumber\\
p_\sigma(0,0) &= (\y_0,\nabla_\f A(\X,\f)), \\
p_\step(\step_e,\sigma_e) &= (\Y,\nabla_\X A(\transport(\step_e)\X,\f+\sigma_e\y)),\nonumber\\
p_\sigma(\step_e,\sigma_e) &= (\y_{\sigma_e},\nabla_\f A(\transport(\step_e)\X,\f+\sigma_e\y)).\nonumber
\end{align}
Solving this system and finding the minimums gives the optimal step lengths \(\hat{\step}_k\) and \(\hat{\sigma}_k\) for the quadratic approximation of the search directions.
For the simultaneous NLCG method the step lengths are then \(\step_k = \hat{\step}_k\) and \(\sigma_k = \hat{\sigma}_k\) while the QN method uses \(\step_k = \beta_\X\hat{\step}_k\) and \(\sigma_k = \beta_\f\hat{\sigma}_k\), where \(\beta_\X, \beta_\f\in\;]0,1]\) are constant underrelaxation parameters.
We then simultaneously update \(\X\) and \(\f\) with \(\X_{k+1}=\transport(\step_k)\X_k\) and \(\f_{k+1} = \f_k+\min(\sigma_\mathrm{M},\sigma_k)\y\) respectively.

We then simultaneously update \(X\) and \(\f\) with \(\X_{k+1}=\transport(\step_k)\X_k\) and \(\f_{k+1} = \f_k+\min(\sigma_\mathrm{M},\sigma_k)\y\) respectively.

\textbf{Remark 2:} The surface~\eqref{eq:surface} is determined by computing \(\nabla_\X A\) and \(\nabla_\f A\) at the trial step. The system of equations~\eqref{eq:surfsystem} could alternatively be determined by computing both \(A\) and \(\nabla_\X A\) or \(A\) and \(\nabla_\f A\) at \((\transport(\step_e)\X,\f+\sigma_e\y)\).

\textbf{Remark 3:} Inclusion of the \(\step\sigma\) cross term would require an extra trial evaluation point for system~\eqref{eq:surfsystem} to be linearly independent.

\section{Numerical experiments}\label{sec:numerics}

We use a two dimensional model problem to compare the sequential and simultaneous NLCG and QN methods.
This model problem is inspired by ensemble DFT, and corresponds to a three dimensional system constrained to two dimensions without spin effects and exchange-correlation terms while taking entropy into account.
The model problem adapted from Reference~\cite{marzari} is
\begin{equation}\label{eq:model_problem}
A(\X,\f) = -\tfrac{1}{2}\mathrm{tr}\big(\X^T\laplace\X\,\diag(\f)\big) + \extpotential^T\density+\tfrac{1}{2} \intpotential^T\density -TS(\f).
\end{equation}
Here \(\laplace\in\Real^{m\times m}\) is the discretized Laplace operator,  \(\extpotential\in\Real^{m}\) the external potential, \(\density \in\Real^m\) the electron density, \(\intpotential = \V\density\) the Hartree potential corresponding to the electron density \(\density\), \(T\) to temperature, and \(S\) is the entropy. The electron density is
\begin{equation}
\density = (\X\circ\X)\,\f,
\end{equation}
where \(\circ\) is the entrywise, or Hadamard, product.
The entropy term is
\begin{equation}
S(\f)=-\sum_{i=1}^n f_i\ln(f_i+\delta(1-f_i))+(1-f_i)\ln(1-f_i+\delta f_i),
\end{equation}
where \(\delta > 0\) is a small regularization parameter that ensures that the derivative of \(S\) remains finite.

To calculate the potentials we use
\begin{equation}
(\extpotential)_i = -\sum_{j=1}^N \frac{Z_j}{\|\position_i - \Position_j\|+\alpha},
\end{equation}
where the sum is over the nuclei with charge \(Z_j\) and position \(\Position_j\). The position corresponding to the discretization point \(i\) is \(\position_i\), and the parameter \(\alpha\) is used to regularize the potential. \(\V\in\Real^{m\times m}\) is similarly given by
\begin{equation}
\V_{ij} = \frac{1}{\|\position_i-\position_j\| + \alpha}.
\end{equation}

We solve the problem in the unit square with zero boundary conditions corresponding to an infinite potential well. We use a uniform finite difference discretization with \(m\) inner points to obtain a system where \(\X \in \Real^{m\times n}\). Here \(n\) corresponds to the number of electronic orbitals in the calculation.
As initial guess we use the solution of the quadratic problem using the first two terms of~(\ref{eq:model_problem}).
The occupation numbers are initialized to
\begin{equation}
f_i = \tfrac{\electrons}{n} + \tfrac{1}{2}\Delta\frac{n+1-2i}{n+1},
\end{equation}
where \(\Delta = \min(\electrons/n,1-\electrons/n)\) and \(\electrons\leq n\) is the number of electrons. This choice ensures that the initial occupation of all orbitals is nonzero and emphasizes lower energy orbitals.

We demonstrate the methods for three external potentials. For all models we use potential regularization \(\alpha = 5\times10^{-2}\) and entropy regularization \(\delta = 10^{-3}\).

The first model is a single nucleus with charge \(Z=2\) centered at the center of the unit square with two electrons. For this system the second and third orbitals are degenerate. We calculate the model with 10 electronic orbitals and a first order finite difference discretization with 25 interior points in one dimension resulting in \(m=625\) spatial degrees of freedom. We will refer to this system as \(Z_2\).

The second model, which we name \(Z_3\)-\(Z_2\), consists of two nuclei, with a nuclei of charge \(Z = 3\) placed at \((\tfrac{1}{3},\tfrac{1}{3})\) and another with charge \(Z=2\) placed at \(\tfrac{2}{3},\tfrac{2}{3}\) and 5 electrons. This system has four well separated electronic orbitals, while the fifth and sixth are relatively close. The computation is initialized with 10 orbitals and 29 interior grid points in one dimension for \(m = 841\).

The last model, \(Z_4\)-\(Z_3\), consists of two nuclei, \(Z = 4\) placed at \((\tfrac{1}{3},\tfrac{1}{3})\), and \(Z=3\) at the grid point closest to \((\tfrac{2}{3},\tfrac{13}{24})\) and 7 electrons. The off diagonal placement is chosen to break the symmetry of the system. This model initially has 14 orbitals and 29 interior grid points in one dimension \((m=841)\).

For the sequential QN orbital minimizer uses the parameters \(\beta_\X = 0.4\), \(\mu = 5\times10^{-5}\), and history length 6. The sequential QN and NLCG methods minimum trial step length \(\step_0 = 10^{-3}\) and we perform 6 orbital optimization steps before engaging the occupation number minimizer. Both sequential optimization routines use an identical SD routine with \(\beta_\f = 0.5\), \(\mu=10^{-4}\) and \(\sigma_0 = 10^{-4}\) for occupation number optimization with two optimization steps.
We have tried several different combinations of orbital and occupation optimization steps and observed that this combination offers a good compromise.
For the SD method \(\mu\) only serves to scale the approximate line search.

We measure convergence by the energy difference to a reference energy computed by running the simultaneous methods for 3000 steps and the sequential methods for 3000 optimization rounds. We then use the lowest energy obtained as the reference energy.

\begin{figure}
\begin{center}
\includegraphics{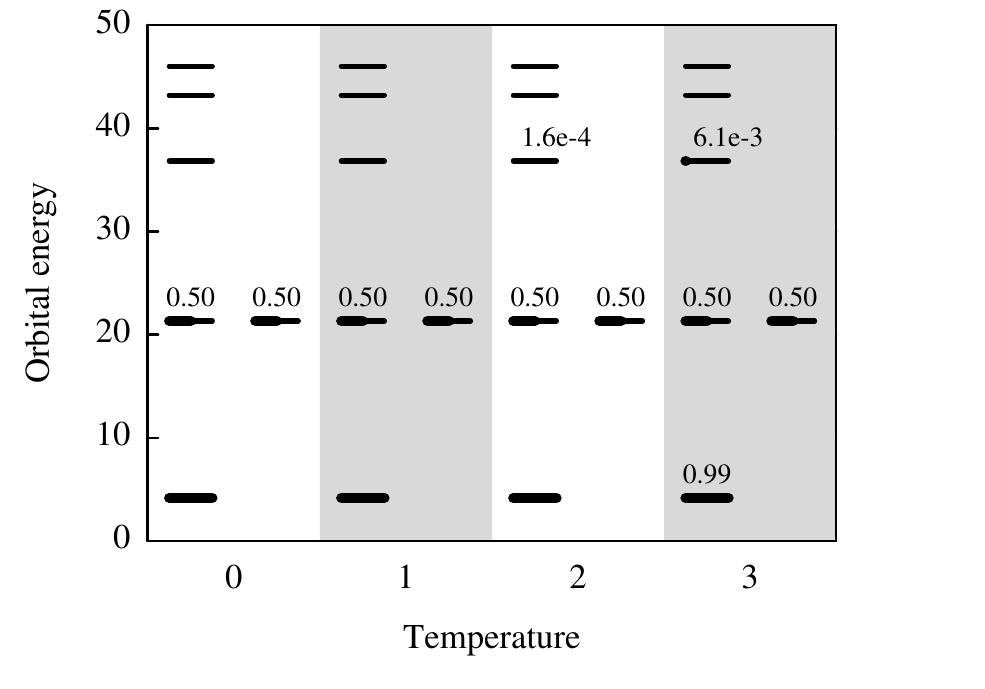}
\caption{\label{fig:temp1}Orbital energy levels with occupation for \(Z_2\) at varying temperatures. Fractional occupation numbers are indicated and the same data is also presented in Table~\ref{tab:z2}.}
\end{center}
\end{figure}
\begin{figure}
\begin{center}
\includegraphics{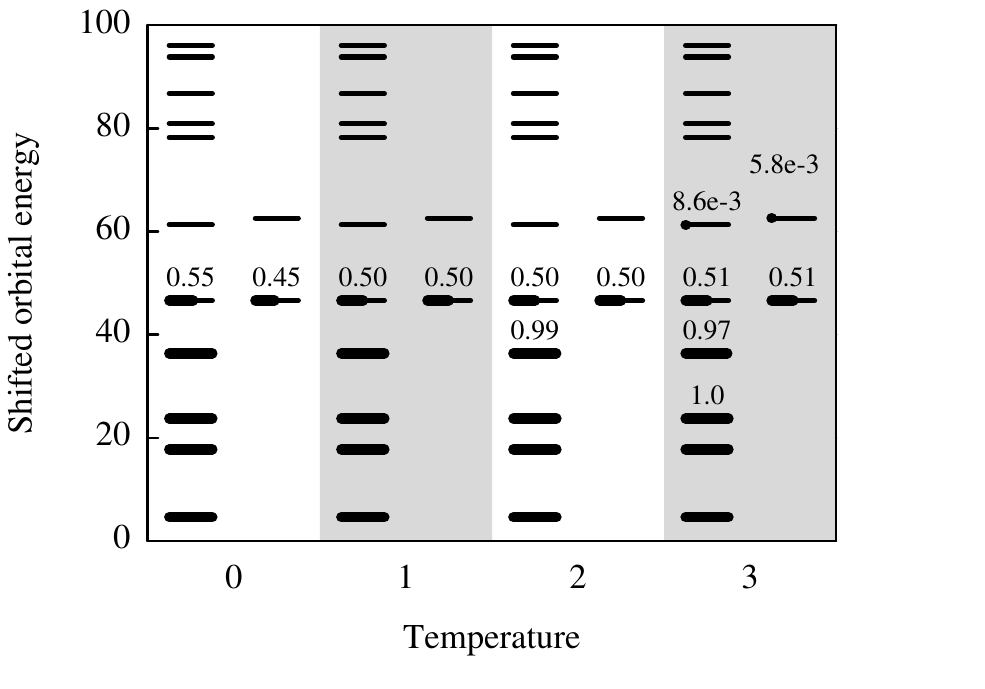}
\caption{\label{fig:temp2}Orbital energy levels with occupation for \(Z_3\)-\(Z_2\) at varying temperatures with orbital energy shifted by \(+5\). Fractional occupation numbers are indicated and the same data is also presented in Table~\ref{tab:z3z2}.}
\end{center}
\end{figure}
\begin{figure}
\begin{center}
\includegraphics{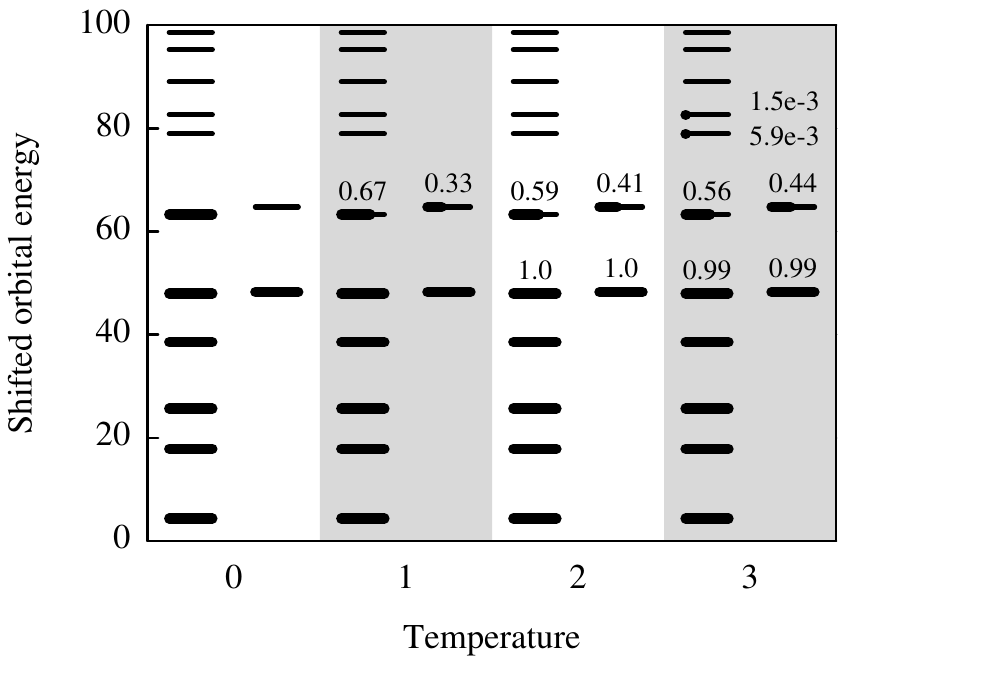}
\caption{\label{fig:temp3}Orbital energy levels with occupation for \(Z_4\)-\(Z_3\) at varying temperatures with orbital energy shifted by \(+10\). Fractional occupation numbers are indicated and the same data is also presented in Table~\ref{tab:z4z3}.}
\end{center}
\end{figure}
The change in occupation numbers with rising temperature is graphically presented in Figures~\ref{fig:temp1},~\ref{fig:temp2}, and~\ref{fig:temp3}.
The same data is repeated in Tables~\ref{tab:z2},~\ref{tab:z3z2}, and~\ref{tab:z4z3}.
At \(T=0\) the lowest electronic orbitals are fully occupied for \(Z_4\)-\(Z_3\), while one electron is split between two degenerate orbitals for \(Z_2\).
Even though there is a small gap (\(1.69\times10^{-2}\)) between the fifth and sixth electron orbitals for \(Z_3\)-\(Z_2\) the fifth electron is split (0.55 vs 0.45) between these orbitals. We successfully replicated this split with a 3000 round sequential SD orbital occupation number optimization. Furthermore, restarting the SD iteration with a five orbital initial guess based on the split orbital reference solution results in convergence to a higher energy state.

\begin{figure}
\begin{center}
\includegraphics{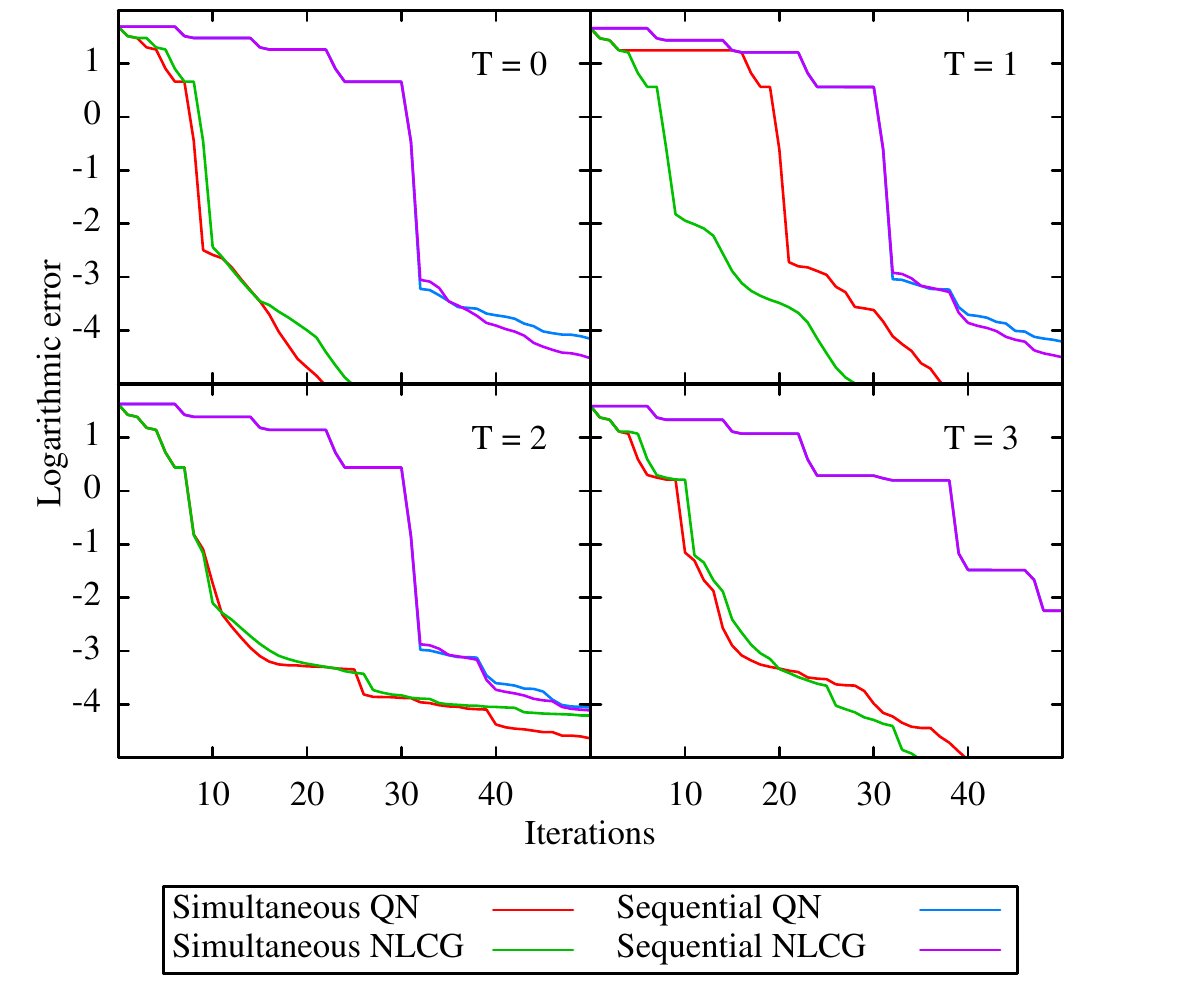}
\caption{\label{fig:system1}Energy convergence for \(Z_2\) at varying temperatures.}
\end{center}
\end{figure}
\begin{figure}
\begin{center}
\includegraphics{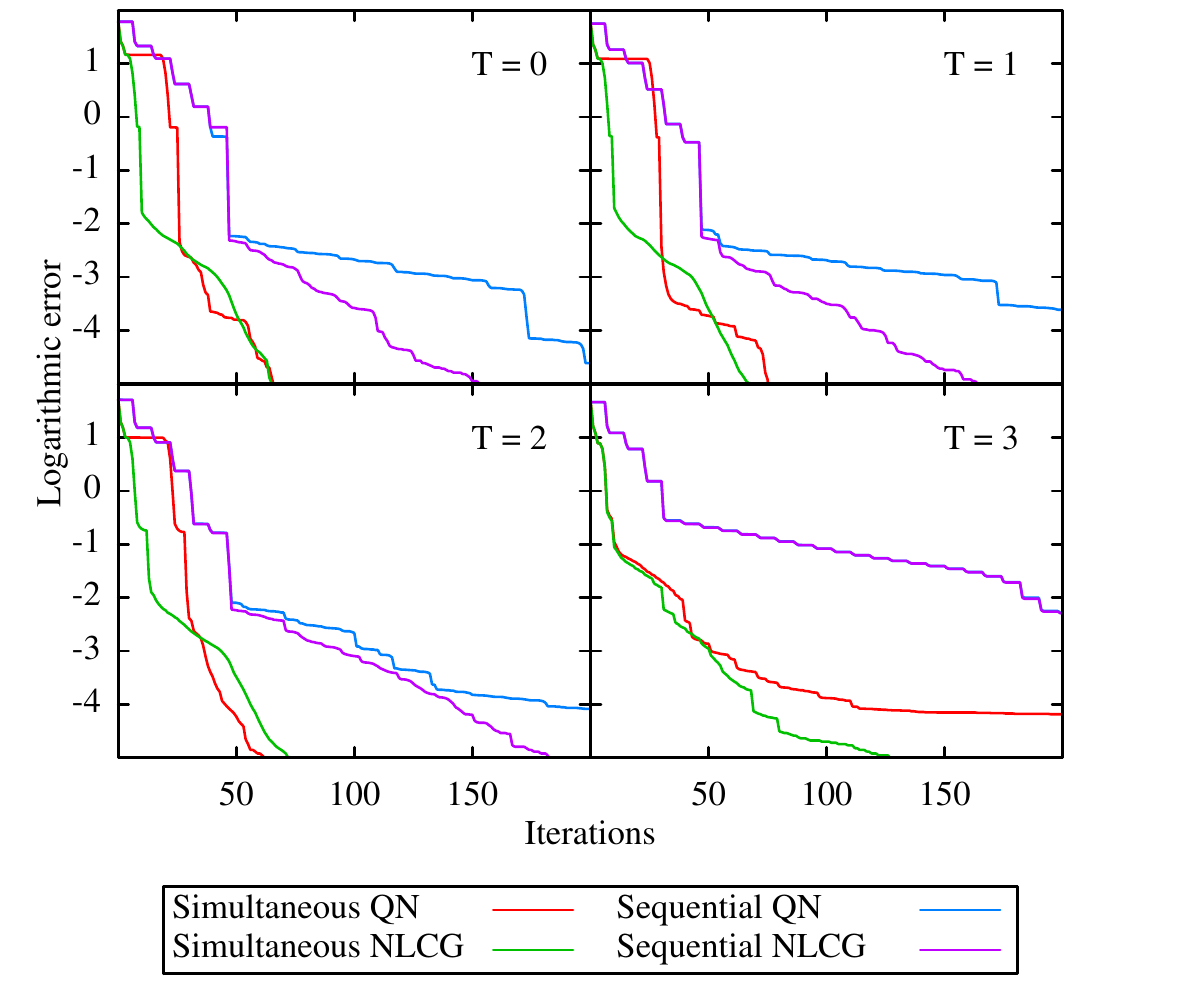}
\caption{\label{fig:system2}Energy convergence for \(Z_3\)-\(Z_2\) at varying temperatures.}
\end{center}
\end{figure}
\begin{figure}
\begin{center}
\includegraphics{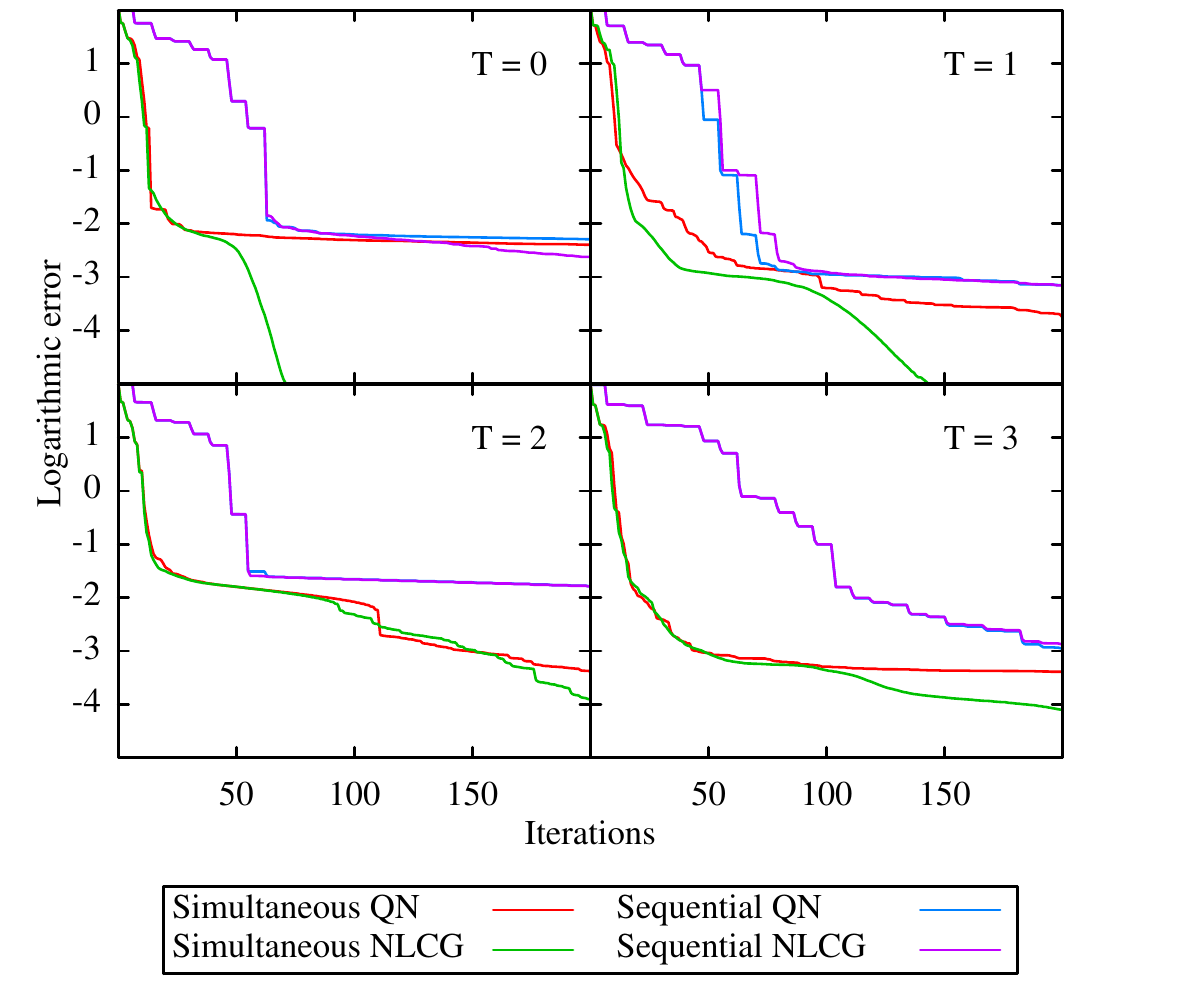}
\caption{\label{fig:system3}Energy convergence for \(Z_4\)-\(Z_3\) at varying temperatures.}
\end{center}
\end{figure}
Figures~\ref{fig:system1},~\ref{fig:system2}, and~\ref{fig:system3} illustrate energy convergence for the different methods.
The simultaneous methods generally perform better than the sequential methods, and the simultaneous NLCG method is more robust than the simultaneous QN approach.
In the energy convergence for the sequential optimization routines the switch between orbital and occupation optimization is readily seen in the steplike energy convergence.
Furthermore, the performance of the sequential QN and NLCG methods is nearly identical for all models. This might be due to the limited number of step available for orbital optimization before occupation optimization is enabled.

\begin{figure}
\begin{center}
\includegraphics{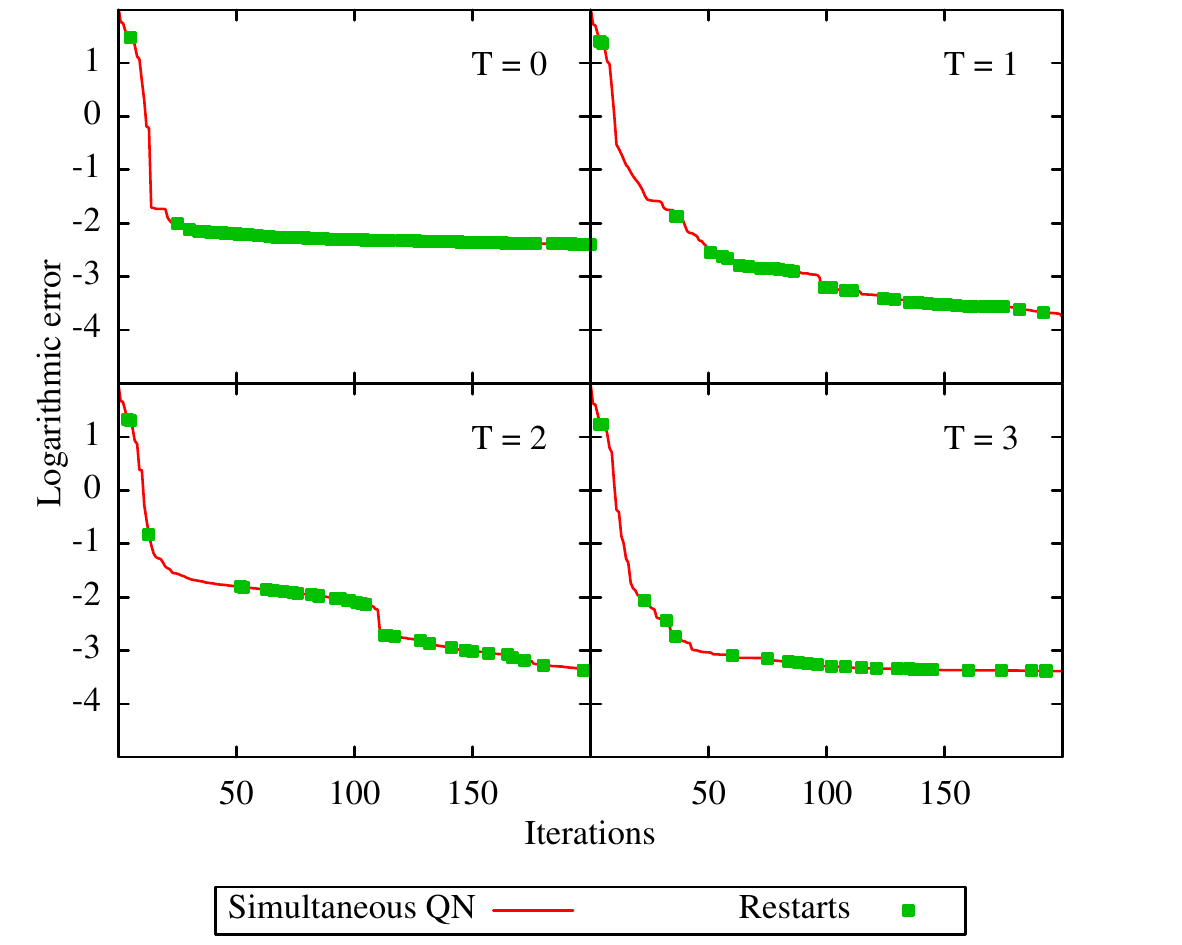}
\caption{\label{fig:system3_restart}Energy convergence of the simultaneous QN method with restarts for \(Z_4\)-\(Z_3\) at varying temperatures.}
\end{center}
\end{figure}
The simultaneous NLCG method outperforms the QN method for the \(Z_4\)-\(Z_3\) system shown in Figure~\ref{fig:system3}. Increasing history generally improves the convergence rate of the QN method, but this did not significantly change the rate of convergence for this model.
Frequent restarts limit history length and provide at least a partial explanation for this effect. Figure~\ref{fig:system3_restart} presents \(Z_4\)-\(Z_3\) restarts for the simultaneous QN method.
Restarts are frequent for this model at all temperatures compared to \(Z_2\) and \(Z_3\)-\(Z_2\).
However, for \(T > 0\) there is generally sufficiently many steps between restarts for the history to grow to full length, and the rate of convergence does improve somewhat.

\begin{figure}
\begin{center}
\includegraphics{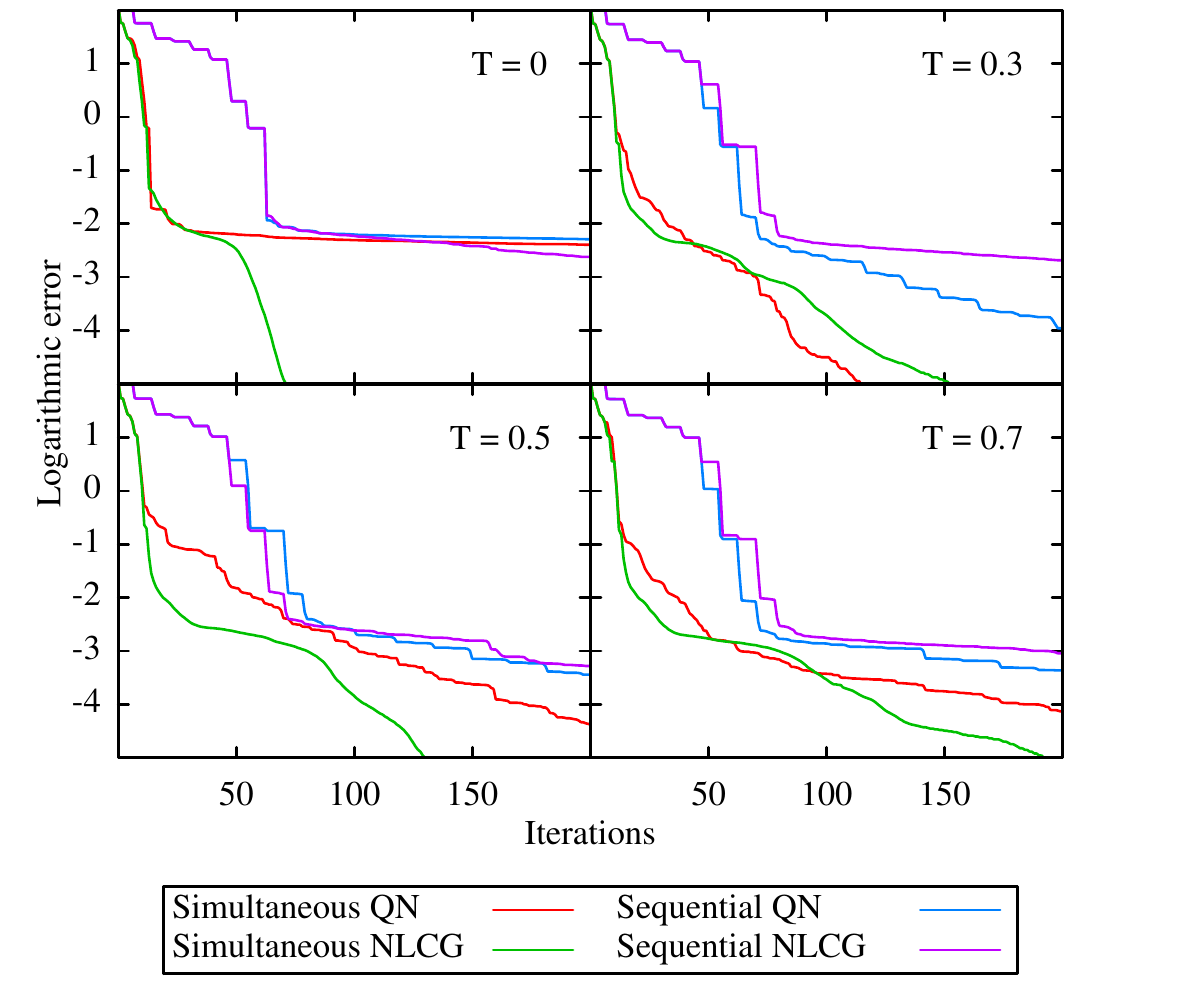}
\caption{\label{fig:system3p}Energy convergence for \(Z_4\)-\(Z_3\) at varying temperatures \(T < 1\).}
\end{center}
\end{figure}
For the \(Z_4\)-\(Z_3\) model the energy difference between the highest occupied and lowest unoccupied orbital is \(1.69\times10^{-2}\), see Figure~\ref{fig:temp3} and Table~\ref{tab:z4z3}.
This difference is comparatively small and could explain the poor performance of the QN method, particularly for \(T=0\).
In Figure~\ref{fig:system3p} the convergence rate of the optimization procedures for \(Z_4\)-\(Z_3\) for \(T = 0.3,0.5,0.7\), and the convergence rate for \(T=0\) is included for reference.
At \(T = 0.3\) the rate of convergence for the QN method is considerably improved and the convergence rate remains superior to \(T=0\) for \(T=0.5\) and \(T=0.7\).
The elevated temperature broadens the Fermi surface and this could explain the improved convergence at \(T=0.3\), while the convergence of higher energy orbitals makes the problem more challenging at higher temperatures.
This would also explain the decreasing performance of NLCG for higher temperatures.

\begin{table}[!hb]
\begin{center}
\caption{\label{tab:z2}Orbital energy levels with occupation for \(Z_2\) at varying temperatures. The same data is graphically presented in Figure~\ref{fig:temp1}.}
\begin{tabular}{ccccc}
\hline
E   &   Occ. (T=0)  &   Occ. (T=1)  &   Occ. (T=2)  &   Occ. (T=3)  \\
\hline
 4.172259  & 1.000000  & 1.000000  & 1.000000  & 0.996380 \\
21.328241  & 0.500000  & 0.500000  & 0.499955  & 0.498751 \\
21.328241  & 0.500000  & 0.500000  & 0.499880  & 0.498751 \\
36.836577  & 0.000000  & 0.000000  & 0.000165  & 0.006117 \\
43.225667  & 0.000000  & 0.000000  & 0.000000  & 0.000000 \\
46.034373  & 0.000000  & 0.000000  & 0.000000  & 0.000000 \\
%60.894571  & 0.000000  & 0.000000  & 0.000000  & 0.000000 \\
%60.894571  & 0.000000  & 0.000000  & 0.000000  & 0.000000 \\
\hline
\end{tabular}
\end{center}
\end{table}

\begin{table}[!hb]
\begin{center}
\caption{\label{tab:z3z2}Orbital energy levels with occupation for \(Z_3\)-\(Z_2\) at varying temperatures with orbital energies shifted by \(+5\). The same data is graphically presented in Figure~\ref{fig:temp2}.}
\begin{tabular}{ccccc}
\hline
E   &   Occ. (T=0)  &   Occ. (T=1)  &   Occ. (T=2)  &   Occ. (T=3)  \\
\hline
 4.606322  & 1.000000  & 1.000000  & 1.000000  & 1.000000 \\
17.773445  & 1.000000  & 1.000000  & 1.000000  & 1.000000 \\
23.744218  & 1.000000  & 1.000000  & 1.000000  & 0.999833 \\
36.378253  & 1.000000  & 1.000000  & 0.994738  & 0.970970 \\
46.607469  & 0.554627  & 0.504114  & 0.504757  & 0.508795 \\
46.624356  & 0.445373  & 0.495886  & 0.500505  & 0.506011 \\
61.308726  & 0.000000  & 0.000000  & 0.000000  & 0.008575 \\
62.566830  & 0.000000  & 0.000000  & 0.000000  & 0.005816 \\
78.212923  & 0.000000  & 0.000000  & 0.000000  & 0.000000 \\
80.870921  & 0.000000  & 0.000000  & 0.000000  & 0.000000 \\
86.752316  & 0.000000  & 0.000000  & 0.000000  & 0.000000 \\
93.853450  & 0.000000  & 0.000000  & 0.000000  & 0.000000 \\
96.049712  & 0.000000  & 0.000000  & 0.000000  & 0.000000 \\
% Unshifted
%-0.393678  & 1.000000  & 1.000000  & 1.000000  & 1.000000 \\
%12.773445  & 1.000000  & 1.000000  & 1.000000  & 1.000000 \\
%18.744218  & 1.000000  & 1.000000  & 1.000000  & 0.999833 \\
%31.378253  & 1.000000  & 1.000000  & 0.994738  & 0.970970 \\
%41.607469  & 0.554627  & 0.504114  & 0.504757  & 0.508795 \\
%41.624356  & 0.445373  & 0.495886  & 0.500505  & 0.506011 \\
%56.308726  & 0.000000  & 0.000000  & 0.000000  & 0.008575 \\
%57.566830  & 0.000000  & 0.000000  & 0.000000  & 0.005816 \\
%73.212923  & 0.000000  & 0.000000  & 0.000000  & 0.000000 \\
%75.870921  & 0.000000  & 0.000000  & 0.000000  & 0.000000 \\
%81.752316  & 0.000000  & 0.000000  & 0.000000  & 0.000000 \\
%88.853450  & 0.000000  & 0.000000  & 0.000000  & 0.000000 \\
%91.049712  & 0.000000  & 0.000000  & 0.000000  & 0.000000 \\
%114.662059  & 0.000000  & 0.000000  & 0.000000  & 0.000000 \\
%115.998571  & 0.000000  & 0.000000  & 0.000000  & 0.000000 \\
%116.869381  & 0.000000  & 0.000000  & 0.000000  & 0.000000 \\
%117.969053  & 0.000000  & 0.000000  & 0.000000  & 0.000000 \\
%131.733688  & 0.000000  & 0.000000  & 0.000000  & 0.000000 \\
%133.968717  & 0.000000  & 0.000000  & 0.000000  & 0.000000 \\
%148.421767  & 0.000000  & 0.000000  & 0.000000  & 0.000000 \\
\hline
\end{tabular}
\end{center}
\end{table}

\begin{table}[!hb]
\begin{center}
\caption{\label{tab:z4z3}Orbital energy levels with occupation for \(Z_4\)-\(Z_3\) at varying temperatures with orbital energies shifted by \(+10\). The same data is graphically presented in Figure~\ref{fig:temp3}.}
\begin{tabular}{ccccc}
\hline
E   &   Occ. (T=0)  &   Occ. (T=1)  &   Occ. (T=2)  &   Occ. (T=3)  \\
\hline
 4.327524  & 1.000000  & 1.000000  & 1.000000  & 1.000000 \\
17.873544  & 1.000000  & 1.000000  & 1.000000  & 1.000000 \\
25.639021  & 1.000000  & 1.000000  & 1.000000  & 1.000000 \\
38.541992  & 1.000000  & 1.000000  & 1.000000  & 1.000000 \\
47.960214  & 1.000000  & 1.000000  & 0.999983  & 0.994464 \\
48.278074  & 1.000000  & 1.000000  & 0.999841  & 0.993917 \\
63.300484  & 1.000000  & 0.669980  & 0.591678  & 0.564432 \\
64.759294  & 0.000000  & 0.330020  & 0.408498  & 0.439697 \\
78.938961  & 0.000000  & 0.000000  & 0.000000  & 0.005937 \\
82.626842  & 0.000000  & 0.000000  & 0.000000  & 0.001553 \\
89.029063  & 0.000000  & 0.000000  & 0.000000  & 0.000000 \\
95.252731  & 0.000000  & 0.000000  & 0.000000  & 0.000000 \\
98.574017  & 0.000000  & 0.000000  & 0.000000  & 0.000000 \\
\hline
\end{tabular}
\end{center}
\end{table}

\section{Conclusion}\label{sec:conclusion}

We have presented two schemes for energy optimization of ensemble DFT computations.
The updates take the problem constraints into account and permits us to use information obtained from previous evaluations of the target functional and gradients to improve rate of convergence. We have further demonstrated the methods numerically on a model problem inspired by the electronic structure theory and compared simultaneous and sequential schemes based on the QN and NLCG methods.

The ensemble model successfully concentrates occupation to low energy orbitals at low temperatures, and gradually increases occupation of higher energy orbitals at increasing temperature to increase the entropy of the system.
Optimization of the occupation numbers also enables ensemble DFT calculations to automatically handle degenerate and near degenerate orbitals at \(T=0\), which are challenging for methods that construct the electron density by the Aufbau principle.
%The optimization procedure also intuitively handles degenerate and near degenerate orbitals at \(T=0\).
Furthermore, is seems possible to broaden the Fermi surface by increasing temperature to accelerate convergence of small gap systems.

Simultaneous optimization schemes provide improved convergence compared to sequential approaches for both the NLCG and QN methods.
While the NLCG and QN methods are often comparable in performance, the NLCG method is overall more robust.
In contrast, Reference~\cite{baarman2011b} found that QN method is more robust than the NLCG method.
It is possible that the quadratic approximate line search gives a better result for the model problem.
As the NLCG method depends heavily on a high quality line search this might provide a possible explanation.
%The QN method performs poorly for problems with frequent restarts and while this effect does not fully explain the lack of convergence it can be used as a problem indicator.
In the present case, the QN method performs poorly for problems with frequent restarts and while this effect does not fully explain the lack of convergence it can be used as a problem indicator.

\newpage
\bibliographystyle{plain}
\bibliography{references}

\end{document}